\documentclass[modern, times]{aastex61}

\usepackage{graphicx}
\usepackage{amsfonts,amsmath,amssymb}
\usepackage[italic]{mathastext}
\usepackage[utf8]{inputenc}

\newcommand\wfirst{\textit{WFIRST}}
\newcommand\gaia{\textit{Gaia}}
\newcommand\hst{\textit{HST}}
\newcommand\webbpsf{{\sc WebbPSF}}

\DeclareRobustCommand{\object}[1]{%
   #1%
}

\begin{document}

\title{In the crosshair: astrometric exoplanet detection with \wfirst's diffraction spikes}

\correspondingauthor{Peter Melchior}
\email{peter.melchior@princeton.edu}

\author{Peter Melchior}
\affiliation{Department of Astrophysical Sciences, Princeton University, Peyton Hall, Princeton, NJ 08544, USA}

\author{David Spergel}
\affiliation{Department of Astrophysical Sciences, Princeton University, Peyton Hall, Princeton, NJ 08544, USA}
\affiliation{Center for Computational Astrophysics, Flatiron Institute, 162 5th Ave, NY, NY 10010}

\author{Arianna Lanz}
\affiliation{Department of Astrophysical Sciences, Princeton University, Peyton Hall, Princeton, NJ 08544, USA}

\shorttitle{Astrometric exoplanet detection with \wfirst}
\shortauthors{Melchior, Spergel, Lanz}

\begin{abstract}
\wfirst\ will conduct a coronagraphic program of characterizing the atmospheres of planets around bright nearby stars. 
When observed with the \wfirst\ Wide Field Camera, these stars will saturate the detector and produce very strong diffraction spikes. 
In this paper, we forecast the astrometric precision that \wfirst\ can achieve  by centering on the diffraction spikes of highly saturated stars.
This measurement principle is strongly facilitated by the \wfirst\ H4RG detectors, which confine excess charges within the potential well of saturated pixels.
By adopting a simplified analytical model of the diffraction spike caused by a single support strut obscuring the telescope aperture, integrated over the \wfirst\ pixel size, we predict the performance of this approach with the Fisher-matrix formalism. 
We discuss the validity of the model and find that $10\,\mu$as astrometric precision is achievable with a single 100\,s exposure of a $R_{AB}=6$ or a  $J_{AB}=5$ star.
We discuss observational limitations from the optical distortion correction and pixel-level artifacts, which need to be calibrated at the level of $10-20\,\mu$as so as to not dominate the error budget.
To suppress those systematics, we suggest a series of short exposures, dithered by at least several hundred pixels, to reach an effective per-visit astrometric precision of better than $10\,\mu$as.
If this can be achieved, a dedicated \wfirst\ GO program will be able to detect Earth-mass exoplanets with orbital periods of $\gtrsim1$\,yr around stars within a few pc as well as Neptune-like planets with shorter periods or around more massive or distant stars.
Such a program will also enable mass measurements of many anticipated direct-imaging exoplanet targets of the \wfirst\ coronagraph and a ``starshade'' occulter.
\end{abstract}


\section{Introduction}

The \wfirst\ mission will use a repurposed 2.4-meter telescope to conduct a program of studying dark energy, detecting planets through microlensing, imaging planets with a coronagraph and using its
wide field camera for general astrophysics \citep{Spergel2013,Spergel2015}.   Since one of the primary science drivers for \wfirst\ is making precision measurements of weak gravitational lensing to characterize 
the nature of dark energy, \wfirst\ is being designed to have a very stable point spread function (PSF): \wfirst\ is operating at the thermally and dynamically stable L2 point and its thermal/mechanical design is optimized to 
minimize variations in the PSF.  \wfirst\ has a wide field camera whose 18 Mercury Cadmium Telluride (HgCdTe) H4RG chips enable a 0.28 square degree field.  The combination of this large field of view that will image hundreds of bright 
\gaia\ \citep{Gaia2016} stars with each observation and the stable PSF makes \wfirst\ a powerful instrument for astrometry.

\wfirst\ should be able to conduct a rich and diverse program of using astrometry to address science questions ranging from the nature of dark matter, to testing stellar models to searching for exosolar planets. 
The \wfirst\ science team reports \citep{Spergel2013,Spergel2015} describe some of these potential applications enabled by \wfirst's ability to obtain submilli-arcsecond astrometry even for stars as faint as 25th magnitude. 
\citet[later G15]{Gould2015} discusses \wfirst\ astrometry as part of the bulge survey.
In this paper, we will focus on the capabilities of \wfirst\ bright star astrometry and its application to detecting exoplanets around nearby stars.   \wfirst's
H4RG HgCdTe  detectors are well suited for bright star astrometry because these CMOS detectors trap charges in pixels, unlike CCDs that bleed.

In this work, we seek to understand how precisely one can determine stellar centroids of very bright stars given the design of \wfirst, foremost its PSF shape and pixel size.
A similar investigation has been performed by G15.
We extend that work in three main aspects:
1) G15 consider only imaging obtained as part of the \wfirst\ microlensing program, which has a fixed exposure time and dither pattern, and only uses the $H$-band. 
We investigate all available filters in the lastest \wfirst\ design and a range of exposure times.
2) Because of the pixel-level artifacts, we specifically want to avoid a strong reliance of the centroid measurement on a small number of pixels, spreading out the signal over a larger area and rejecting the inner regions of the PSF even if they are not saturated.
Our results are therefore more conservative than those of G15, who found that a sizeable fraction of the statistical power stems from mildly saturated pixels in the core.
3) We attempt to determine the amount of uncorrected systematics and propagate these uncertainties to the final precision.
By splitting a visit into several exposures, we explore the trade-offs between exposure time and systematics mitigation to achieve optimal per-visit astrometric precision.

Because we intentionally avoid the core regions, we can restrict our model to capture only the features that are relevant at high flux levels, namely the diffraction spikes, and generate an analytic model that can be evaluated in all bands (\S\ref{sec:model}) and estimate the sensitivity of that model to shifts in the centroid of the star with the Fisher-matrix approach (\S\ref{sec:fitting}).
In \S\ref{sec:limitations} we discuss how various observational effects limit the centroiding precision, and investigate in \S\ref{sec:exoplanets} if \wfirst\ could successfully detect exoplanets from diffraction-spike measurements.
We conclude in \S\ref{sec:conclusion}.

Throughout this work, all magnitudes are in the AB system unless otherwise noted.

\section{The PSF model}
\label{sec:model}

\begin{figure*}[t]
  \includegraphics[width=0.33\linewidth]{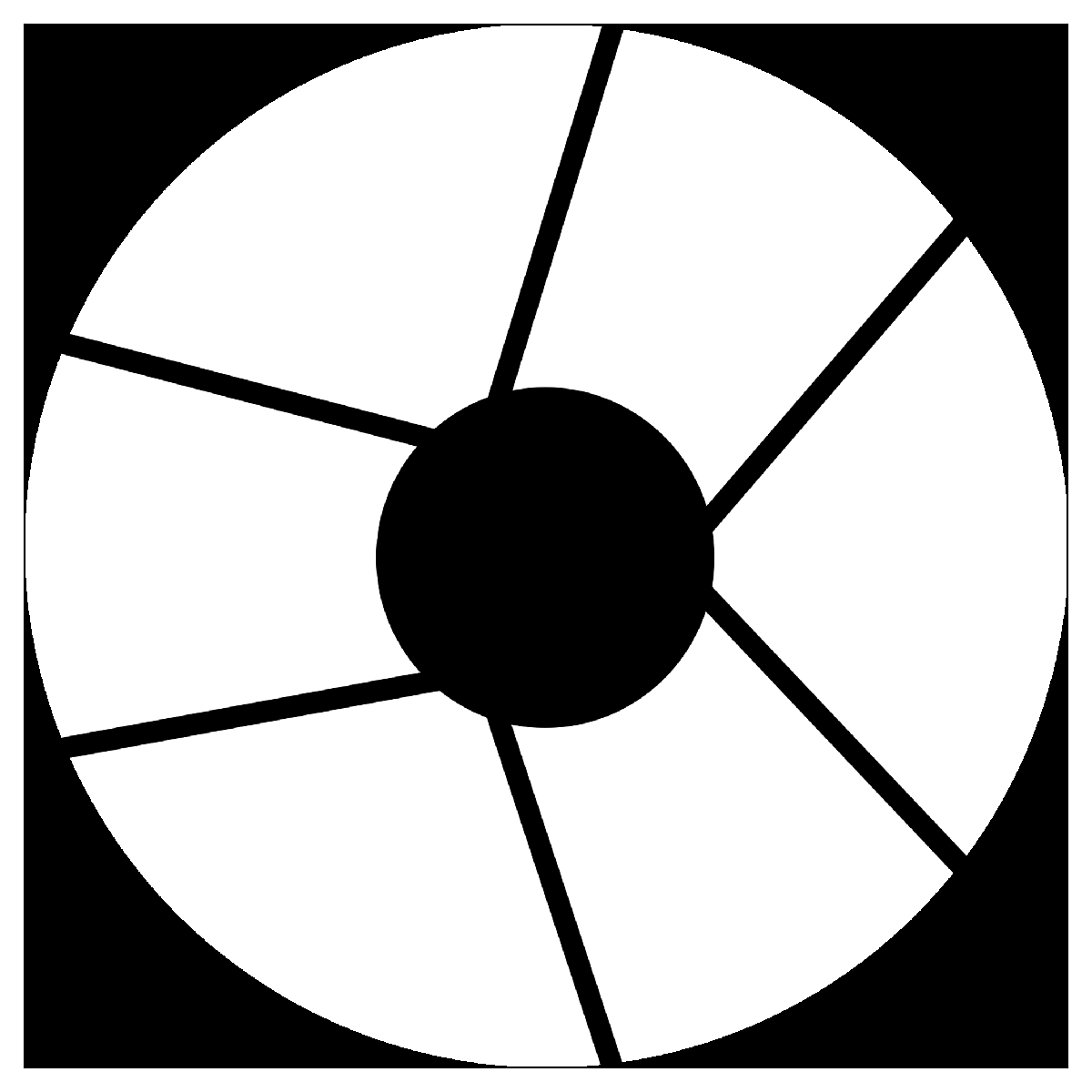}
   \includegraphics[width=0.33\linewidth]{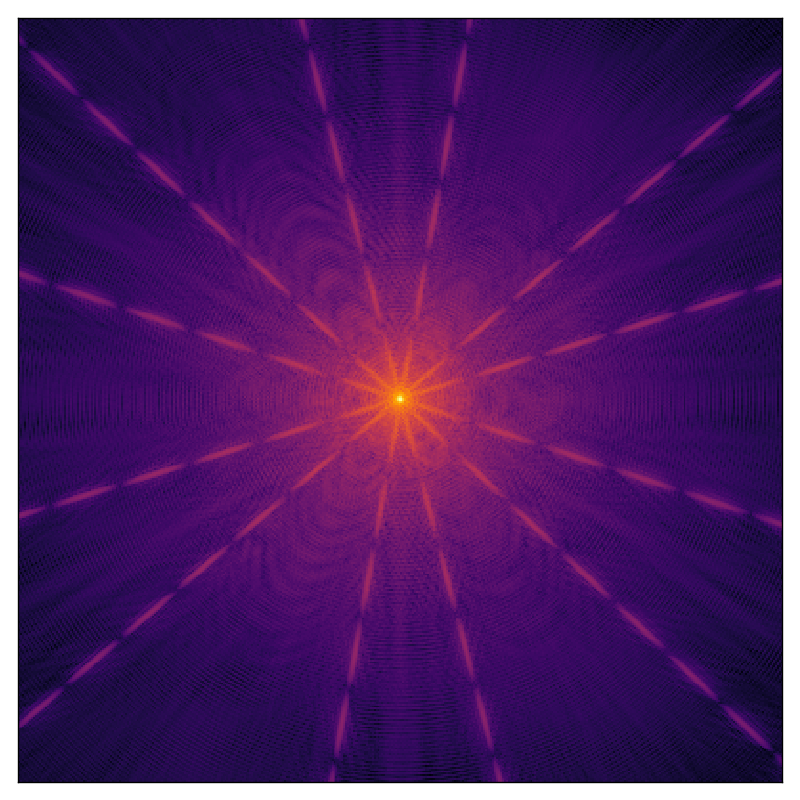}
   \includegraphics[width=0.33\linewidth]{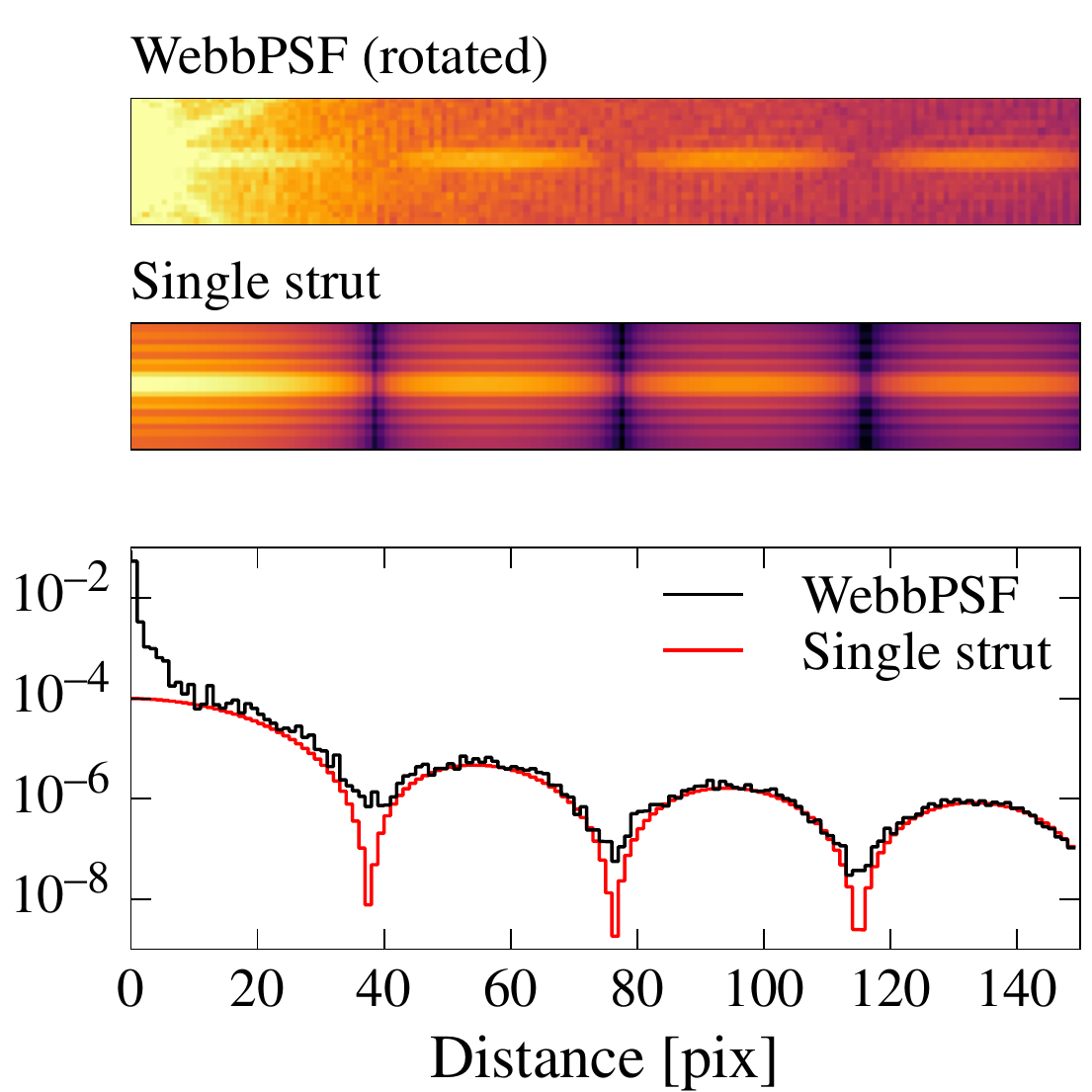}
  \caption{\emph{Left:} \wfirst\ WFC Cycle 5 pupil for filters from R062 to H158. For reference, the aperture diameter is 2.37\,m. \emph{Center:} \webbpsf\ model of a monochromatic point source with $\lambda=1\,\mu$m in the center of the focal plane. Colors have logarithmic stretch. The presence of 12 diffraction spikes instead of 6 is a consequence of the non-radial alignment of the support struts. \emph{Right:} Visual comparison and horizontal profiles at peak intensity of the  \webbpsf\ model and the analytical model of a single support strut given in \autoref{eq:psf2d}. The \webbpsf\ model was internally oversampled by a factor 10, rotated, and then downsampled to the final resolution.}
\label{fig:wfirst_pupil_psf}
\end{figure*}

We will approximate the shape of the PSF along a diffraction spike as being purely caused by a single support strut of the secondary mirror obscuring a part of the telescope aperture.
In addition, we assume that the strut is rectangular with sidelengths $a$ and $b$, pointing in radial direction from the center to the edge of the pupil.
The actual \wfirst\ pupil deviates from those assumptions: there is an inner and outer aperture radius, and the struts are not exactly radially aligned (see left panel of \autoref{fig:wfirst_pupil_psf}).%
\footnote{For observations well above 1\,$\mu$m, an additional pupil mask is mounted directly on the filter, increasing the obscured areas of the pupil and noticeably changing the PSF and diffraction spike shapes, rendering our model inapplicable. Given the currently planned telescope operating temperature of 260 K, the pupil mask would only be used for the filters redward of  H158, which we will therefore neglect in this study.}
We will nonetheless adopt this simplified pupil and investigate the accuracy of the model later.

In the Fraunhofer regime of geometrical optics, we can describe the electrical field $E(x,y)$ at the focal plane in angular coordinates as the Fourier transform of the rectangular obstruction:
\begin{equation}
\label{eq:E}
E(x,y) = \text{sinc}\left(\frac{\pi a \sin(x)}{\lambda}\right)\text{sinc}\left(\frac{\pi b \sin(y)}{\lambda}\right)
\end{equation}
The PSF intensity is given by the square of the electric field,
\begin{equation}
\label{eq:psf}
I(x,y) = \text{sinc}^2\left(k_x x\right)\text{sinc}^2\left(k_y y\right)\text{,}
\end{equation}
where we employed the small-angle approximation and introduced $k_x \equiv \pi a\big/\lambda$ and $k_y \equiv \pi b\big/\lambda$. 
This well known result is not directly applicable because the diffraction features are narrow compared to the pixel grid.
We therefore need to account for the pixelation, i.e. each \wfirst\ pixel has a finite size $w=$ 0.11 as, which amounts to an integration over the box-shaped pixel area,
\begin{equation}
\label{eq:psfp}
I_p(x,y) = \int dx'dy' I(x',y')\ \text{rect}\left(\frac{x-x'}{w}\right) \text{rect}\left(\frac{y-y'}{w}\right)
\end{equation}
followed by sampling that function at the centers of a pixel grid enumerated with indices $(i,j)$,
\begin{equation}
\label{eq:sampling}
I_p(i,j) = I_p\big(x_i, y_j),
\end{equation}
where $x_i = (i+\tfrac{1}{2})w$.
Given the form of \autoref{eq:psfp}, the resulting function must be separable,
\begin{equation}
\label{eq:psf2d}
I_p(x,y\,|\,k_x,k_y,w) = I_p(x\,|\,k_x,w) I_p(y\,|\,k_y,w).\\
\end{equation}
The analytical form of the one-dimensional $I_p(x)$ is given in \autoref{eq:psf1d}.
A comparison between this PSF model of a single support strut and the \webbpsf\footnote{\url{http://www.stsci.edu/wfirst/software/webbpsf}} model \citep{Perrin2012,Perrin2014} of the entire pupil  is show in right panel of \autoref{fig:wfirst_pupil_psf}.

\begin{figure}[t]
  \centering\includegraphics[width=0.8\linewidth]{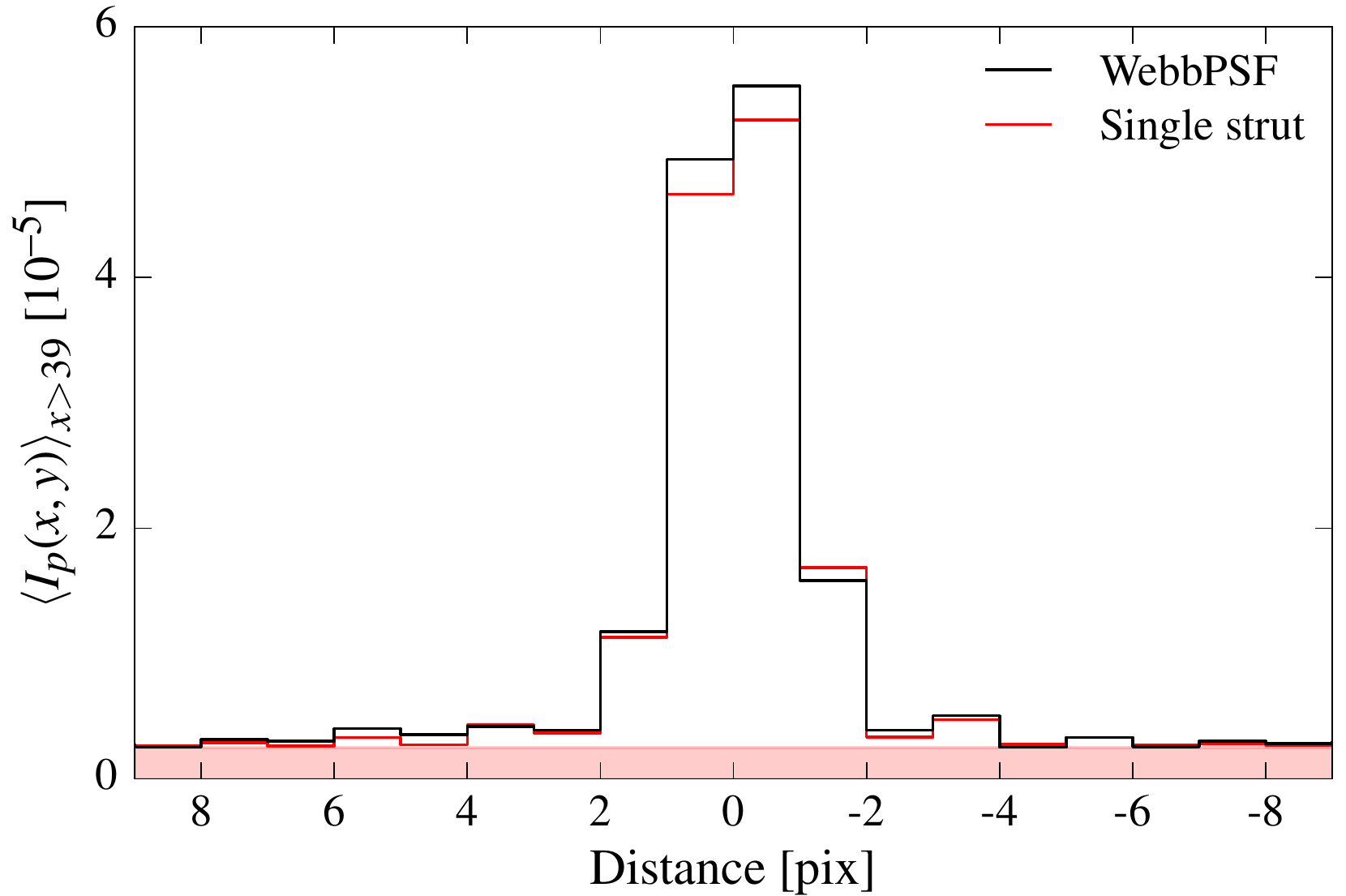}
  \caption{Comparison of the vertical profile of the diffraction spikes as predicted by \webbpsf\ and the analytical model of a single support strut given in \autoref{eq:psf2d} for a point source with $\lambda=1\,\mu$m in the center of the focal plane. The area within the first diffraction minimum (here $x<39$\, pix) was excluded. The model underestimates the amount of diffracted light from the centrally obscured pupil, which can be approximated by a constant in $y$-direction with an amplitude of $0.2\cdot10^{-5}$ (shown in \emph{light red}).}
\label{fig:psf_y}
\end{figure}

We can see that the analytical model correctly captures the main features of the diffraction spike, in particular the location of the minima in $x$-direction, from which we determine $a=4.85$\,cm.
The profile in $y$-direction, which contains almost all information about the stellar centroid, is shown in \autoref{fig:psf_y}, from which we infer $b=84$\,cm.
Both values of $a$ and $b$ are in good agreement with measurements from the pupil image shown in the left panel of \autoref{fig:wfirst_pupil_psf}.

It is worthwhile noting that the oscillations of the diffractions spikes are resolved with the \wfirst\ pixel scale $w$, even at $\lambda<1\,\mu$m, while the core of the PSF is undersampled.
This is a consequence of the smaller length $b$ of the support strut compared to the full aperture diameter.

On the other hand, we can also see the limitations of the analytical model: 1) It drastically underestimates the intensity in the center. 
We will therefore restrict the fitting range to outside of the first minimum in $x$-direction, i.e. $|x| > \lambda\big/a$, which corresponds to 39 pix at $\lambda=1\,\mu$m.
2) There is a floor of diffracted light from the centrally obscured pupil (see central panel of \autoref{fig:wfirst_pupil_psf}), which reduces the  dynamic range in the actual diffraction spike, especially in $y$-direction.
At fixed $x$, it can be approximated by a constant, but it reduces the contrast of the higher-order diffraction features.
We will therefore restrict the fitting range to the inner 3 maxima, i.e. $|y| < 3 \lambda\big/b$, or $6$ pix at $\lambda=1\,\mu$m.

\section{Centering on the diffraction spikes}
\label{sec:fitting}

We make use of the Fisher information to determine the precision with which we can fit for the centroid given the pattern of the diffraction spike in data $D$,
\begin{equation}
F_{i,j} = -\frac{\partial^2}{\partial\theta_i \, \partial\theta_j} \ln \mathcal{L}(D;\theta)\mathrm{,}
\end{equation}
with two fit parameters $\theta=(x,y)$.
We assume Gaussian and uncorrelated noise\footnote{Correlated noise, e.g. from remapping the images onto an undistorted frame, can be considered by introducing a pixel covariance matrix in the following equations. We neglect it in this work because we expect correlated noise not to significantly alter our findings.}, which leads to the familiar $\chi^2$ form
\begin{equation}
\label{eq:lnL}
\ln \mathcal{L} = -\frac{1}{2}\sum_{i,j\in\mathcal{V}}\frac{\left[D_{i,j} - N f_\mathrm{strut} I_p(x_i - x, y_j - y)\right]^2}{\sigma_{i,j}^2},
\end{equation}
where the normalization accounts for the total flux that is incident on the strut: $N$ is the total number of photons to reach the unobscured telescope aperture, and $f_\mathrm{strut}=ab\big/[(1-o)(d/2)^2\pi]$ is the fraction of the aperture covered by the strut. 
According to the WFIRST Cycle 6 telescope parameters, the mirror diameter $d=2.37$\,m, and the central obscuration $o$ corresponds to 13.9\%, resulting in $f_\mathrm{strut}\approx 0.01$.

The likelihood should only be evaluated for valid pixels $(i,j)\in\mathcal{V}$, which combines two conditions: the range in which the PSF model is a fair description of the actual diffraction spike (see \S\ref{sec:model}), and an intensity limit set by the saturation level of the H4RG detectors, estimated at $I_p(i,j)<1.2\cdot10^5\,\mathrm{e}^-$ (B. Rauscher, private comm.) with a gain of 2.

Because of the separability of $I_p$ we can express the elements of the Fisher matrix as
\begin{equation}
\label{eq:Fisher}
\begin{split}
F_{11} &= N^2 f^2_\mathrm{strut} \sum_{i,j\in\mathcal{V}} \frac{I_p(y_j-y)^2}{\sigma_{i,j}^2}\left[\frac{\partial I_p(x_i - x)}{\partial x}\right]^2\\
F_{22} &= N^2 f^2_\mathrm{strut}\sum_{i,j\in\mathcal{V}} \frac{I_p(x_i-x)^2}{\sigma_{i,j}^2}\left[\frac{\partial I_p(y_j - y)}{\partial y}\right]^2\\
F_{12} &= N^2 f^2_\mathrm{strut}\sum_{i,j\in\mathcal{V}} \frac{I_p(x_i-x, y_j-y)}{\sigma_{i,j}^2} \frac{\partial I_p(x_i - x)}{\partial x} \frac{\partial I_p(y_j - y)}{\partial y}.
\end{split}
\end{equation}
The one-dimensional derivatives are listed in \autoref{eq:dpsf1d}. 

The final ingredient for the likelihood are the pixel noise variances, for which we assume a combination of Poisson, sky background, thermal ermission, read-out noise, and dark current
\begin{equation}
\label{eq:noise}
\sigma_{i,j}^2 = N f_\mathrm{strut} I_p(i,j) + I_\mathrm{sky} + I_\mathrm{thermal} + \sigma^2_\mathrm{ron} + \sigma^2_\mathrm{dark}
\end{equation}
The read-out noise and dark currents are not yet known ($\sigma_\mathrm{ron}<20\, \mathrm{e}^-\,\mathrm{pix}^{-1}$ as per \wfirst\ Science Requirements). We adopt $\sigma_\mathrm{ron}= 5\,\mathrm{e}^-\,\mathrm{pix}^{-1}$ and $\sigma_\mathrm{dark} = 0.015\, \mathrm{e}^- \mathrm{s}^{-1} \mathrm{pix}^{-1}$.
Thermal and sky background intensities, taken from the \wfirst\ exposure time calculator\footnote{\url{https://wfirst.ipac.caltech.edu/sims/tools/wfDepc/wfDepc.html}}, for different wavelengths are listed in \autoref{tab:noise}.

\begin{table}
\caption{Thermal and sky background intensities in the \wfirst\ filter bands, with a spacecraft operating temperature of 260 K and filter throughputs set to 0.95. The filter names denote the central wavelength, e.g. Y106 implies $\lambda_\mathrm{c}=1.06\,\mu$m.}
\label{tab:noise}
\begin{tabular}{lcc}
\hline\hline
Filter & $I_\mathrm{sky}\ [\mathrm{e}^- \mathrm{s}^{-1} \mathrm{pix}^{-1}]$ & $I_\mathrm{thermal}\ [\mathrm{e}^- \mathrm{s}^{-1} \mathrm{pix}^{-1}]$\\
\hline
R062 & 0.638 & 0.023\\
Z087 & 0.464 & 0.023\\
Y106 & 0.453 & 0.023\\
J126 & 0.442 & 0.023\\
H158 & 0.437 & 0.052\\
\hline
\end{tabular}
\end{table}

The relevant quantity in this configuration is the uncertainty in the narrow $y$-direction,
\begin{equation}
\Delta_y^2 = \left[F^{-1}\right]_{22},
\end{equation}
where we marginalized over the uncertainty in the $x$-direction.
However, \wfirst\ has $k=6$ support struts at different angles $\phi_k$, separated by approximately 30 deg, so that each of them can be used as an independent measurement of $y$ with precision $\Delta_y\cos(\phi_k)$, resulting in a joint astrometric precision of either $x$ or $y$ of
\begin{equation}
\label{eq:delta_sum} 
\Delta_\mathrm{pos} \approx \sqrt{\frac{2}{6} \left[F^{-1}\right]_{22}}.
\end{equation}
Note that a more precise measurement could be made by exploiting the fact that each diffraction spike carries information on $x$ \emph{and} $y$, modulated by $\phi_k$, but the improvement is only of order $a\big/b\approx5\%$ compared to \autoref{eq:delta_sum}.

\begin{figure}[t]
\centering\includegraphics[width=0.8\linewidth]{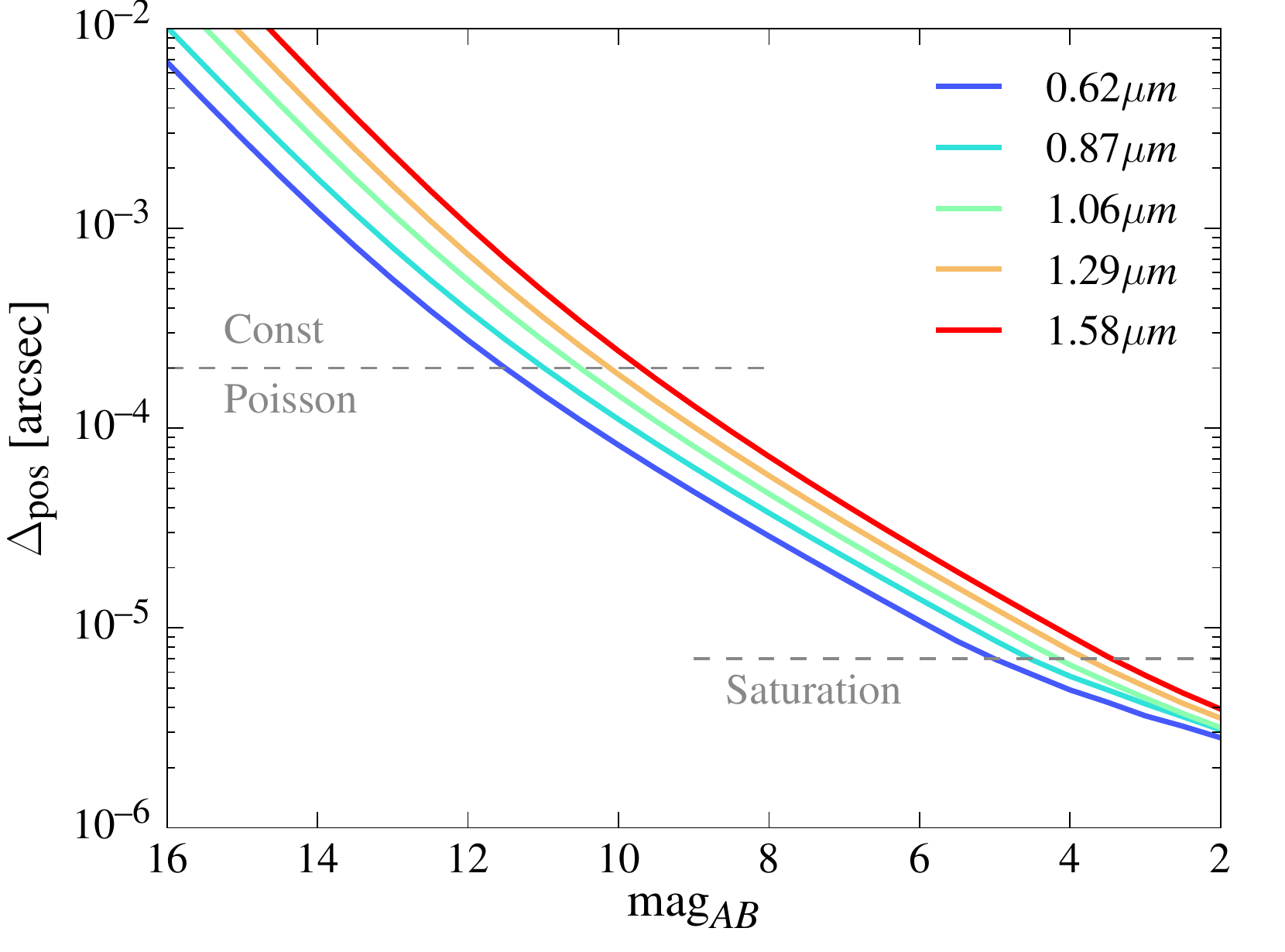}
\caption{Astrometric precision as a function of stellar magnitude at the given wavelength for an exposure time of 100\,s.
The precision was averaged over 30 randomly chosen subpixel positions of the stellar center.
The approximate locations for transitions between regimes that determine the shape of each curve are shown as horizontal dashed lines. 
Best precisions are achieved for the shortest wavelengths because the diffraction spike is narrowest.}
\label{fig:precision}
\end{figure}

In \autoref{fig:precision} we show the predicted precision for a $t=100$\,s exposure as a function of stellar magnitude.
The model is calculated for a monochromatic star with a wavelength centered on the \wfirst\ filters, and the conversion between photon count and magnitude incorporates Cycle 6 specifications of the optical transmission, detector efficiency, and contamination losses.
 
For fixed wavelength, the shape of the curve is dominated, from left to right:
1) by the constant, i.e. source-independent noise terms, foremost the sky background;
2) by the Poisson noise of the diffraction spike itself, scaling with $N^{-\frac{1}{2}}$; 
3) by saturation, which impedes further improvements once the pixels in the region with a valid PSF model start to saturate, removing them from $\mathcal{V}$. 
We can see that for most of the magnitude range, the shortest wavelength yields the best astrometric precision because the diffraction spike is narrowest and sky and thermal emissions are lowest (cf. \autoref{tab:noise}).
But at short wavelengths the steep profiles of the diffraction spikes saturate more quickly, even outside of the first diffraction minimum.

We conclude that the centering on the diffraction spikes of \wfirst\ allows for an astrometric precision of 10\,$\mu$as (equivalent to $\approx10^{-4}$ of a pixel) for a $R=6$ star in a 100 s exposure.
Our estimates in \autoref{fig:precision} agree well with \citet{Gould2015}, who find $\Delta_\mathrm{pos}\approx 10\,\mu$as for a $H_\mathrm{Vega}=3$ (approximately $H=4.4$) star in a 52\,s exposure, for which they utilized both the diffraction spikes and mildly saturated pixels.
We want to stress that this combination is useful because mildly saturated pixels, which become saturated only after the first non-destructive read(s) of an exposure, provide a noticeable amount of astrometric information.
The gain comes from utilizing pixels \emph{between} the diffraction spikes, which requires an accurate PSF model for all pixels in question.
Our simplified analytical model does not consider these areas and still captures the main aspects of the astrometric measurement in the highly saturated regime.

\section{Limitations}
\label{sec:limitations}

So far we have calculated the astrometric precision one could achieve if the location of each pixel on the sky were perfectly known and reproducible, particularly important for long-running campaigns to establish variations in the stellar locations.
Several effects will limit the precision in practice.

\subsection{Optical distortion correction}

As all optical instruments, \wfirst\ will exhibit geometric distortions that need to be corrected when image positions are mapped to locations on the sky.
The precision of this mapping depends on the availability of an ideally distortion-free reference catalog, for instance from \gaia, or the ability to self-calibrate the distortions by constructing an internal master catalog of celestial object positions \citep[e.g.][]{Anderson2003}.
We expect the self-calibration approach to yield superior results because \wfirst\ would be able to utilize stars (and possibly galaxies) below the magnitude limit of \gaia\ and would not have to extrapolate the apparent motions from the \gaia\ reference frame to the epoch of observations years later.
Instead, one is either restricted to an instantaneous astrometric frame or needs to determine parallaxes and peculiar motions of suitable stars from repeated observations separated by years.

The approach has been employed with \hst\ \citep{Anderson2003,Bellini2011} and ground-based imagers \citep{Libralato2014}, with astrometric precisions after correction of the order of 0.1 -- 1~mas.
This is at least an order of magnitude larger than what we seek to achieve for exoplanet detection, but the design and observation strategy of \wfirst\ should enable higher precision because of the thermally stable environment in L2 orbit and the abundance of calibration products from the microlensing survey.
A detailed assessment of the optical distortion correction will be presented by Bellini et al.~(in prep.). 
Here we adopt the following, admittedly optimistic, assumptions:
a template library of optical distortion patterns and an accurate PSF model for each exposure, approximately 10,000 unsaturated stars and compact galaxies per exposure, and 10 back-to-back exposures of approximately 100~s integration time, dithered by hundreds of pixels to uniquely determine the distortion template, should yield an astrometric precision in the central regions of each SCA of 10 $\mu$as.

\subsection{Pixel-level effects}

In addition to residuals of the distortion correction, each pixel will be slightly offset from its assumed location in the focal plane.
Several known detector effects are responsible for such shifts.

The sensitivity of the pixels is not strictly uniform.
\citet{Barron2007} and \citet{Hardy2014} demonstrated that HgCdTe detectors exhibit subpixel quantum efficiency (QE) variations of several percent in some pixels.
While \wfirst\ will be equipped with more advanced H4RG detectors, spatial offsets of a few percent of the pixel width need to be expected (M. Shao, private comm.).
If the subpixel QE variations are uncorrelated between neighboring pixels, the per-pixel offset will be averaged over all pixels in \autoref{eq:lnL}, but most damage would be done in those pixels with a high photon count.
Considering \autoref{fig:wfirst_pupil_psf} and \autoref{fig:psf_y}, most of the flux is accumulated in a narrow strip of approximately $4\times120$ pix, of which the inner $4\times40$ pix are masked.
Over all 12 half-spikes, that amounts to 3840 pix.
An uncorrected spatial offset of 1\%$w$ would then be averaged down to 18 $\mu$as, which demonstrates the benefit of spreading out the signal along the diffraction spikes.
However, for stars with $R<7$ or $H<5$ subpixel QE variation would dominate the error budget.
We anticipate that those variations can be calibrated with laboratory tests on the ground and in flight by the \wfirst\ microlensing survey \citep[their section 2.5.7]{Spergel2015} so that they will not limit the measurements proposed here.

Another relevant effect is nonlinearity, especially the so-called ``brighter-fatter'' effect, which presents itself as an increase in the width of point-sources as a function of source flux.
It can be interpreted as the shrinking of the pixel depletion region due to the charges that accumulate during integration \citep[and references therein]{Plazas2017}.
As the total number of charges is expected to be conserved, the proposed solutions entail a re-apportioning of the pixels fluxes based on the brightness of the source across neighboring pixels.
We expect that such a correction will be sufficient for our purposes.
As long as the exposure times are fixed, any remaining residual of the correction would result in a very subtle additional blur that affects all exposures equally and does therefore not lead to additional astrometric residuals.
It may, however, cause a mild degradation of the pixel-to-pixel contrast in the narrow direction and thus reduce the statistical power of the measurement.

Finally, we address persistence, which denotes a slowly fading imprint of bright objects after an exposure has been read out or even after the telescope has been moved.
Should any preceding exposure induce persistence in a localized region, then the astrometry inferred from areas with a strong flux gradient in the responsible exposure will be biased.
However, the affected regions are entirely predictable, and we can thus slew the telescope to avoid them, or mask them afterwards.

\subsection{Additional effects}

During the exposures the telescope may experience jitter and roll angle changes.
Jitter is the solid-body motion caused predominantly by structural resonances induced by the reaction wheels of the attitude control system.
It is likely that the occurrence of jitter can be diagnosed from telemetry data of that system, but even then its frequency of order 10 Hz will lead to an additional blurring of the entire exposure with an constant direction and an amplitude of up to 14 mas \citep{Spergel2015}.  
As the diffraction spikes are measured contemporaneously with the field stars, that blurring will reduce the centering sensitivity in the jitter direction but not induce an astrometric shift of the target star with respect to the stars that determine the astrometric solution.
This remains true as long as the telescope responds as a solid body; if additional movements like ``beam walk'' are excited, they will have to be diagnosed and corrected from the optical distortion patters they create.  

Rotations of the space craft, specifically roll angle changes, can be diagnosed from the location of stars in subsequent up-the-ramp samples of the exposure.
Instantaneous roll angle changes \citep[their section 2.4.6]{Riess2014} may require additional telemetry data for a reliable determination of the time-averaged roll angle during the exposure.
A model of the PSF model should then generated for each up-the-ramp samples to reflect the corresponding roll angles.  

Finally, in our calculations we have assumed the point source to be monochromatic, while real stars have a continuous spectrum.
That does not affect our findings.
We anticipate that the wavelength dependence of the PSF will be well characterized as part of the \wfirst\ High-Latitude Survey weak-lensing program, and that the spectrum of most stars that are sufficiently bright to serve as targets will be known, so that we can calculate the PSF model for each such star.

\section{Exoplanet detection with \wfirst}
\label{sec:exoplanets}

\subsection{Per-visit astrometry}
Given the limitations we discussed in \S\ref{sec:limitations}, it appears most beneficial to split a single visit of a target star into $E$ exposures with integration times $t_\mathrm{e}$ chosen such that the astrometric precision from \S\ref{sec:fitting} is comparable to the systematics errors.
If those errors are uncorrelated, the overall positional error is
\begin{equation}
\label{eq:delta_visit}
\Delta_\mathrm{pos,v} = \frac{1}{\sqrt{E}} \sqrt{\Delta^2_\mathrm{pos}(t_\mathrm{e}\,|\,\mathrm{mag}, \lambda) + \Delta_\mathrm{sys}^2},
\end{equation}
where $\Delta^2_\mathrm{pos}$ is given by \autoref{eq:delta_sum}.
We adopt a fiducial value $\Delta_\mathrm{sys}=20\,\mu$as, noting that is could be different by factors of several.
To avoid spatial correlations of pixel-level artifacts and to aid the optical distortion correction, we need to slew the telescope between exposures by $\gtrsim200\,$pix, which incurs a slew-and-settle time $t_\mathrm{ss} \approx 20$\,s (J. Kruk, private comm.),
resulting in a total visit time 
\begin{equation}
t_\mathrm{v} = E \,t_\mathrm{e} + (E-1)\,t_\mathrm{ss}.
\end{equation}
Fixing $t_\mathrm{v}$ and varying $E$ determines the best per-visit error of the program. For a reliable optical distortion correction we assume that $E \geq 10$ and $t_\mathrm{e} \geq 10\,$s.

\begin{figure}[t]
\centering\includegraphics[width=0.8\linewidth]{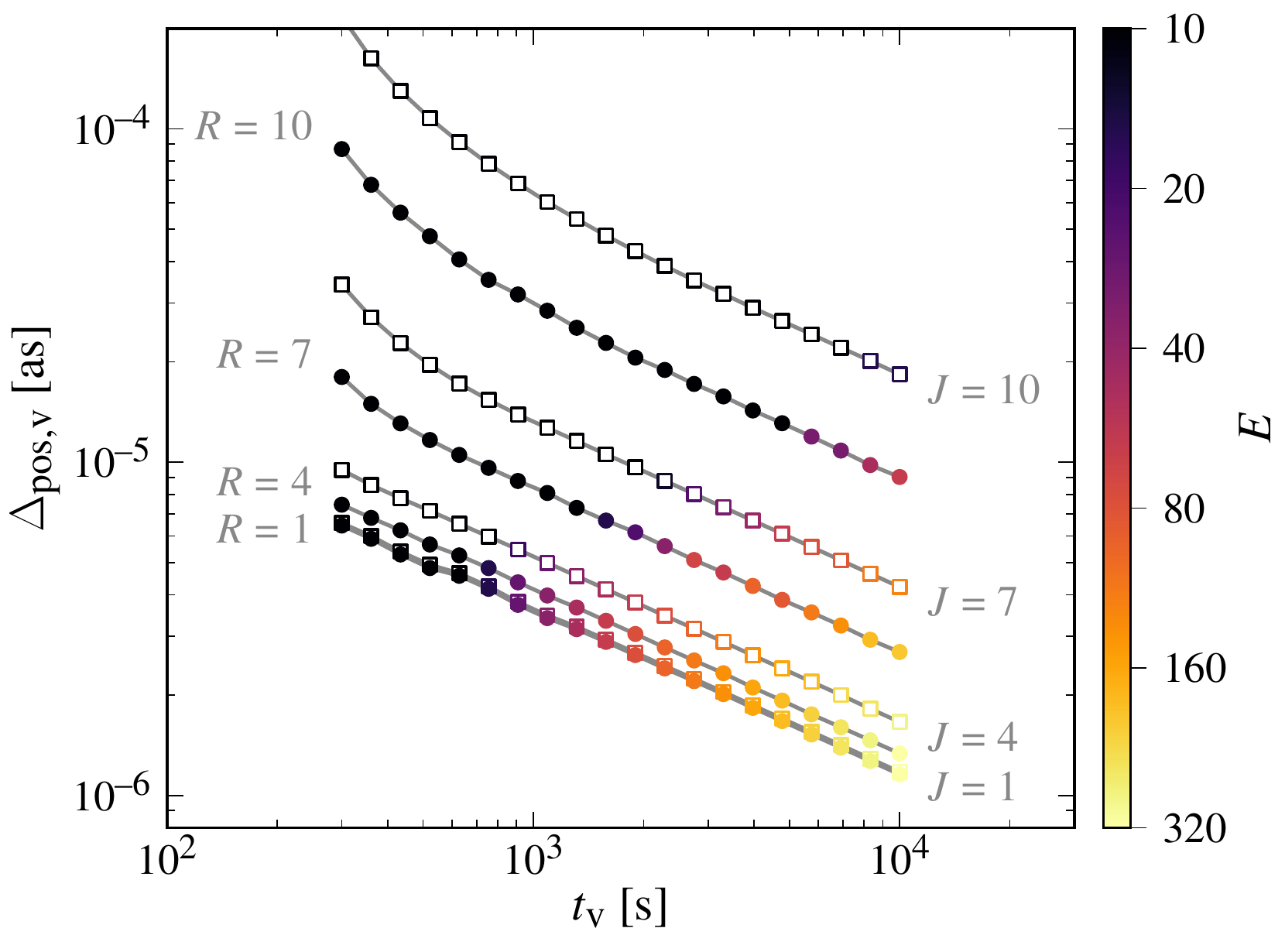}
\caption{Per-visit astrometric precision as a function of integration time per visit $t_\mathrm{v}$ for $\lambda=0.62\,\mu$m ($R$, \emph{full circles}) and $\lambda=1.06\,\mu$m ($J$, \emph{open squares}).
Shown are the best achievable precisions after optimizing over the number of exposures $E$, indicated by the color of the marker.
Different lines correspond to the indicated stellar magnitude in the respective bands.
Faint stars mandate few but long exposures to suppress Poisson noise, for bright stars a large number of exposures is preferable to suppress systematic errors.}
\label{fig:visit}
\end{figure}

The results are shown in \autoref{fig:visit} for two central wavelengths and a range of stellar brightnesses and visit times, starting at the minimum of 280\,s permitted under those constrains.
Several aspects are remarkable.
1) As we have seen in \S\ref{sec:fitting}, the astrometric precision at the same apparent brightness is better in $R$ than in $J$.
2) For stars brighter than $R\approx 7$ or $J\approx7$, the scaling is entirely dependent on the systematic error term in \autoref{eq:delta_visit}, which is minimized by increasing the number of exposures $E$, even though a large fraction of the visit time is spent slewing. 
3) For stars brighter than $R\approx4$ or $J\approx 4$, the magnitude-dependent error term becomes irrelevant, there are thus no substantial gains in astrometric precision.
4) Positional uncertainties of $\Delta_\mathrm{pos,v} \lesssim 10\,\mu$as can be achieved for $R<7.5$ or $J<6.5$ stars with visit times below $1000\,$s.

Besides the assumption of $\Delta_\mathrm{sys}=20\,\mu$as, which sets the overall systematics floor, at which only increasing $E$ provides any gains in accuracy, two other aspects of this forecast are worth pointing out. 
The minimum exposure time $t_\mathrm{e} \geq 10\,$s is rather short for obtaining a sufficiently reliable optical distortion correction, which depends on precise measurements of reference stars. 
If it needs to be increased, the number of exposures $E$ at fixed visit time would have to be reduced accordingly, reducing overall precision for bright stars.
Fainter stars would generally benefit from longer integration times, which at short $t_\mathrm{v}$ can only be realized by reducing $E$.
If the optical distortion correction is stable and a reference catalog can be constructed from all exposures of successive visits, while solving for the apparent motion of the stars between visits, one could reduce the number of per-visit exposures, resulting in increased precision for faint stars.
In summary, the results in \autoref{fig:visit} may be optimistic for bright stars and pessimistic for fainter ones.

\subsection{Exoplanet detectability}

Given that we can achieve astrometric precisions of $10\,\mu$as or better with sufficiently long visit times, detecting exoplanets around nearby stars becomes feasible (e.g. \citealt{Perryman2014,Sozzetti2015} for studies on \gaia's capabilities).
We now seek to determine the general characteristics of exoplanet systems detectable with diffraction-spike measurements.

The astrometric signature of an exoplanet of mass $M_\mathrm{p}$, orbiting a star of mass $M_*\gg M_\mathrm{p}$ with a semi-major axis $a$ at a distance of $d$ from the observer is given by
\begin{equation}
\left(\frac{\alpha}{1\,\mathrm{as}}\right) = \frac{M_\mathrm{p}}{M_*} \left(\frac{a}{1\,\mathrm{AU}}\right) \left(\frac{d}{1\,\mathrm{pc}}\right)^{-1}.
\end{equation}
With Kepler's third law,
\begin{equation}
p^2 = \frac{4\pi^2}{G\,M_*} a^3,
\end{equation}
we can relate $a$ to the orbital period $p$ and re-express the astrometric signature in convenient units:
\begin{equation}
\left(\frac{\alpha}{1\,\mathrm{as}}\right) = \frac{M_\mathrm{p}}{M_\odot} \left(\frac{p}{1\,\mathrm{yr}}\right)^\frac{2}{3}\left(\frac{M_*}{M_\odot}\right)^{-\frac{2}{3}}\left(\frac{d}{1\,\mathrm{pc}}\right)^{-1}.
\end{equation}
For reference, the astrometric signature of the Earth-Sun system at a distance of 1\,pc is $\approx3\,\mu$as.
With the measurement precision attainable here, exoplanet detection prefers planets that are either more massive or on longer period orbits than Earth or stars less massive than the Sun.

\begin{figure}[t]
\centering\includegraphics[width=0.8\linewidth]{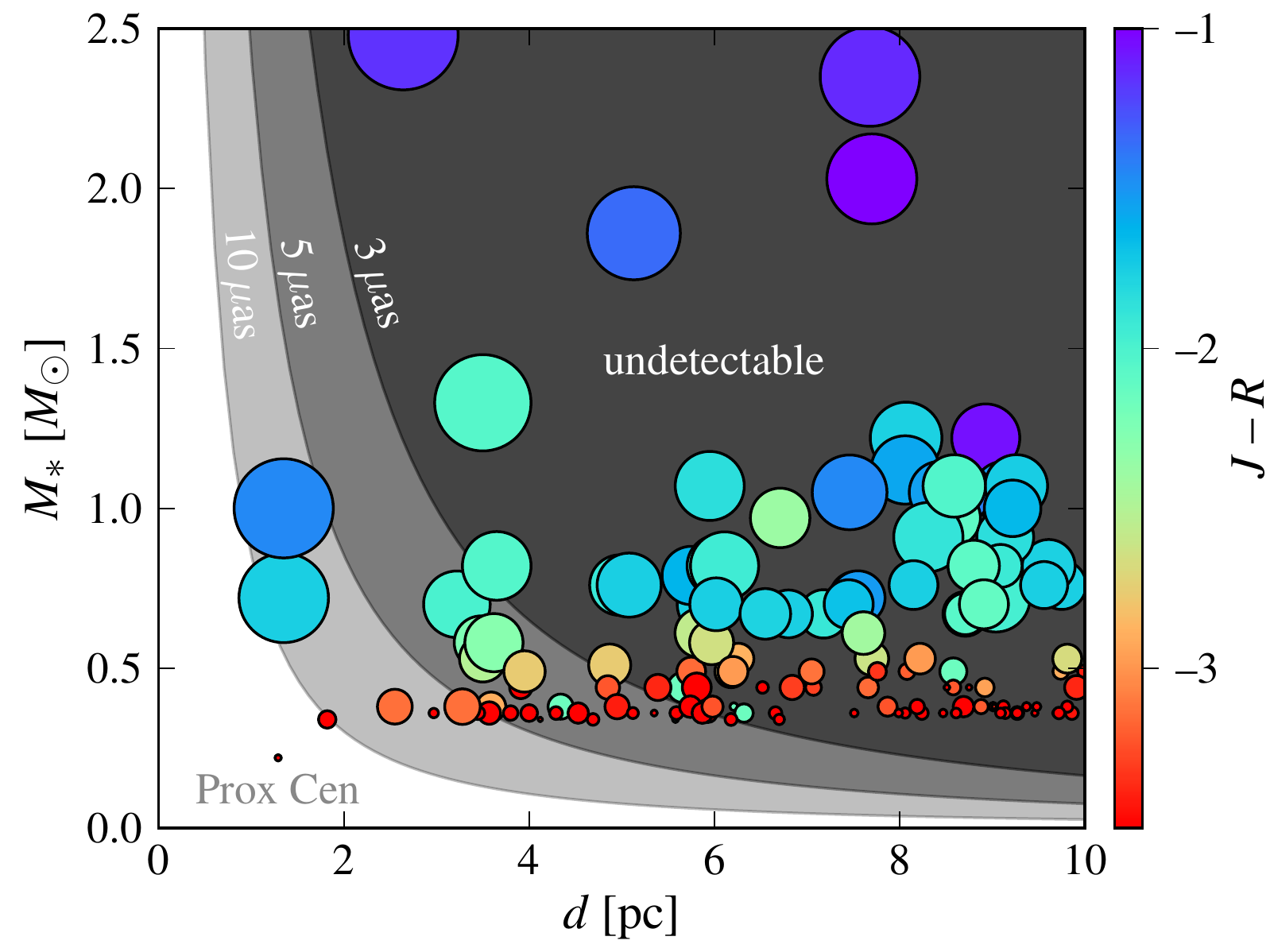}
\caption{Stars from the Extended Hipparcos Compilation within $d=10\,$pc and $V_\mathrm{Hip} < 11$. 
Masses are calculated from the spectral type. 
Colors indicate $J-R$, sizes the apparent brightness in $V_\mathrm{Hip}$. 
The shaded regions correspond to detections given astrometric signatures of $\lbrace3,5,10\rbrace\,\mu$as for a hypothetical planet with $M_\mathrm{p} = 3M_\oplus$ and a period of 1\,yr (lighter is easier to detect).
}
\label{fig:stars}
\end{figure}

The Extended Hipparcos Compilation \citep[XHIP]{Anderson2012} lists 141 stars within 10\,pc with magnitudes of $V_\mathrm{Hip} < 11$, which is approximately the range of stars useful for such a program. 
We take magnitudes $R_J$ and $J_\mathrm{Vega}$ from XHIP and transform them to the AB system according to \citet{Frei1994} and \citep{Blanton2005}.%
\footnote{For a stars on our list, XHIP does not list $R_J$ or $J_\mathrm{Vega}$. We then estimates the magnitudes from $V_\mathrm{Hip}$ and the spectral type.}
We estimate their mass from the spectral type, and show their distance--mass distribution in \autoref{fig:stars}.
We also show the range of stellar masses and distances that would correspond to an astrometric signature of $\lbrace3,5,10\rbrace\,\mu$as for a hypothetical planet with $M_\mathrm{p} = 3M_\oplus$ and a period of 1\,yr.

\begin{figure}[t]
\centering\includegraphics[width=0.8\linewidth]{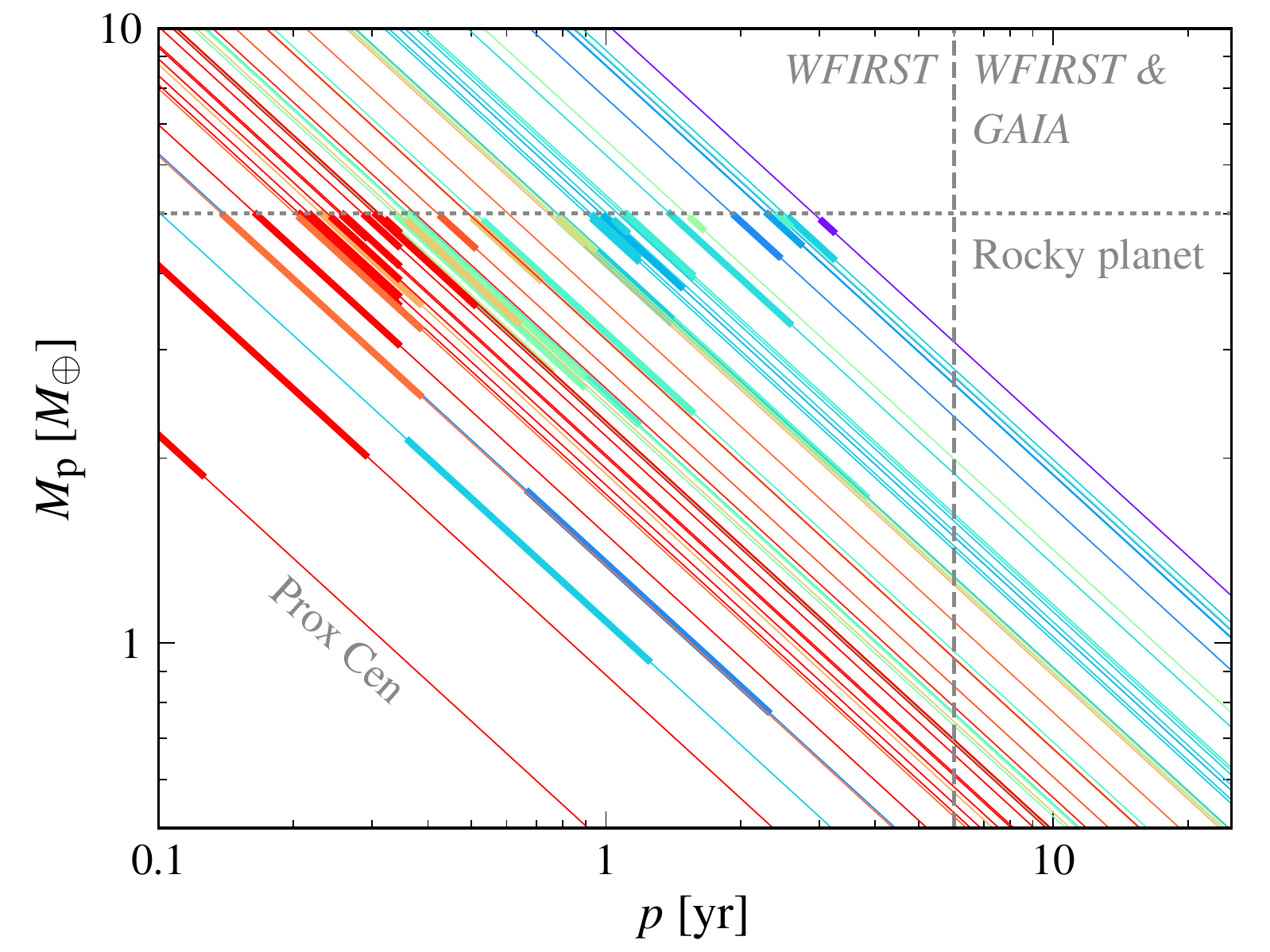}
\caption{Minimal detectable mass of hypothetical planets around the most promising target stars from \autoref{fig:stars}, assuming an astrometric precision of 3\,$\mu$as.
The color scheme is identical to \autoref{fig:stars}.
Thick parts of the lines indicate the habitable zone of the star for a rocky planet (applicable for $M\mathrm{p}<5M_\oplus$, see text for details).}
\label{fig:mass-period}
\end{figure}

The minimal detectable planet mass for the best target stars are shown in \autoref{fig:mass-period}; the ten stars with the best astrometric precisions are listed in \autoref{tab:targets}.
We assume that the stars host a single-planet system and a detection threshold of $\Delta_\mathrm{pos,v}=3\,\mu$as.%
\footnote{This detection criterion is obviously simplistic. In detail, the ability to detect an exoplanet sensitively depends on the orbital parameters, the number and configuration of planets in the system, as well as several observational parameters besides the per-visit astrometric precision, including cadence and total duration of the program \citep{Casertano2008, Perryman2014}. A thorough investigation of \wfirst's exoplanet detection is thus beyond the scope of this work.}
While we adopt the most powerful prediction of the per-visit astrometry for this figure, this is still somewhat conservative because detection should be possible with several visits even if the planet has a smaller $\alpha$. 
Earth-mass planets become detectable around several of the nearby stars, albeit with orbits often longer than 1\,yr.
As expected, dwarf stars provide the best targets, irrespective of their distance.
It is therefore not surprising that \object{Proxima Centauri} (\object{$\alpha$ Cen C}) is by far the best individual target given its low mass and distance of only 1.29\,pc, followed by \object{Gliese 699}, and the two other, more massive stars of the \object{$\alpha$ Cen} system.

We also investigate whether the hypothetical planets would occupy the habitable zones (HZ) of their stars. 
Using the optimistic limits (``recent Venus'' to ``early Mars'' from \citealt{Kopparapu2013}) for planets with rocky composition ($M\mathrm{p}<5M_\oplus$, \citealt{Seager2010}), which depend only on stellar flux and temperature, we determine the corresponding periods $p$ and show them with thick lines in \autoref{fig:mass-period}.
Stars with effective temperatures below 2600\,K or above 7200\,K are excluded, and we ignore any dependence on the planet mass, which mostly affects the inner edge of the HZ \citep{Kopparapu2014}.
Unfortunately, low-mass stars, nominally best for astrometric measurements, have HZs with small $p$, a regime that is hard to access astrometrically.
Nonetheless, with an astrometric precision of $3\,\mu$as, we would  be able to find planets with rocky compositions at the outer edges of their HZs.
In addition, the $G$- and $K$-type stars \object{$\alpha$ Cen A} and \object{$\alpha$ Cen B} are the only stars, for which Earth-mass planets could be found with Earth-like periods. 

Several more stars from our list (not shown in \autoref{fig:mass-period} for the sake of clarity) would add discovery potential of Neptune-class planets, especially on long orbits. 
Realistically, long orbital periods are limited by the lifetime of \wfirst, currently specified as 6\,yr \citep{Spergel2015}. 
A substantial extension to $\approx16\,$yr is possible if we add measurements from \gaia.

\begin{table}[t]
\caption{Target starts with the highest astrometric precision from the Extended Hipparcos Compilation within $d=10\,$pc and $V_\mathrm{Hip} < 11$ (cf. \autoref{fig:stars}).
We list spectral type, distance, and $V_\mathrm{Hip}$ directly from XHIP, ordered by a astrometric sensitivity $\eta$, which we define minimum mass of a single planet on a circular orbit with a period of 1\,yr assuming an astrometric precision of 3\,$\mu$as.}
\label{tab:targets}
\begin{tabular}{lccccc}
\hline\hline
Name & Type & $d$ [pc] & $V_\mathrm{Hip}$ & $\eta$ [$M_\oplus$]\\
\hline
$\alpha$ Cen C & M6 & 1.29 & 10.76 & 0.47 \\
Gliese 699 & M4 & 1.82 & 9.49 & 0.89\\
$\alpha$ Cen B & K1 & 1.35 & 1.24 & 1.08\\
Gliese 411 & M2 & 2.55 & 7.51 & 1.34\\
$\alpha$ Cen A & G2 & 1.35 & 0.14 & 1.35\\
Gliese 729 & M3 & 2.97 & 10.41 & 1.50\\
Gliese 887 & M2 & 3.28 & 7.42 & 1.72\\
Gliese 725 B & M3 & 3.45 & 10.00 & 1.75\\
Gliese 725 A & M3 & 3.57 & 8.92 & 1.81\\
Gliese 15 A & M2 & 3.59 & 8.15 & 1.88\\
\hline
\end{tabular}
\end{table}

\subsection{Synergies with other instruments}

\wfirst's coronagraphic instrument (CGI)  should be the most sensitive coronagraph for the foreseeable future with the ability to achieve contrast ratios of 10$^9$, a thousand-fold increase over {\it JWST} and {\it HST}.  With an inner working angle of $\sim 0.1$\,as, it will be targeting many exoplanet systems that are also favorable astrometric targets: planets around bright stars with periods of $\sim 1-10$ years.  These targets will range from super-Earths to Jupiters.  There are several potential synergies between \wfirst\ astrometric and coronagraphic observations.  The astrometric program could discover planets that are potential targets for coronagraphic observations.  Most exciting, by observing the same planet with the coronagraph and a program of astrometric observations, we will have measurements of the planet's mass and the composition of its atmosphere.

\wfirst\ is also being built to be ready for a ``starshade'' occulter.  The combination of a starshade  and \wfirst\ will enable higher throughputs and contrast ratios than a coronagraph \citep{Roberge2015}.  Just as with the CGI, the astrometric measurements are a powerful complement to the exoplanet imaging.  The astrometry can make the imaging observations more efficient and can increase the confidence of a detection.  Again, the combination of a mass measurement and atmospheric characterization will yield deeper insights into exoplanet properties.

\autoref{tab:synergies} lists some of the anticipated  targets of the \wfirst\ CGI and starshade (J. Kasdin, B. Nemati; private comm.).
In the last column, we list the astrometric sensitivity $\eta$ (same definition as in \autoref{tab:targets}) to illustrate minimal detectable planet masses.
Given that the listed starshade targets are more nearby, astrometric measurements will be more sensitive overall, however, there is a large overlap with Neptune-like planets that CGI is expected to target.
For stars with larger distances, astrometric targets would be restricted to Jupiter-like planets, again in good complementarity to the coronagraph.

\begin{table}[t]
\caption{Prospective targets for the \wfirst\ CGI and starshade. We list spectral type, distance, and $V$ magnitude, and astrometric sensitivity $\eta$, which we define minimum mass of a single planet on a circular orbit with a period of 1\,yr assuming an astrometric precision of 3\,$\mu$as.}
\label{tab:synergies}
\setlength{\tabcolsep}{.25em}
\begin{tabular}{lcccccc}
\hline\hline
Name & CGI & Starshade & Type & $d$ [pc] & $V$ & $\eta$ [$M_\oplus$]\\
\hline
$\alpha$ Cen A & & $\checkmark$ & G2 & 1.35 & 0.14 & 1.35\\
$\alpha$ Cen B & & $\checkmark$ & K1 & 1.35 & 1.24 & 1.08\\
$\alpha$ CMa A & & $\checkmark$ & A1 & 2.64 & -1.47 & 4.21\\
$\epsilon$ Eri & $\checkmark$ & $\checkmark$ & K2 & 3.22 & 3.87 & 2.54\\
61 Cyg & & $\checkmark$ & K5 & 3.48 & 5.37 & 2.42\\
$\alpha$ CMi A & & $\checkmark$ & F5 IV-V & 3.51 & 0.34 & 4.60\\
$\epsilon$ Ind & & $\checkmark$ & K5 & 3.62 & 4.83 & 2.52\\
$\tau$ Cet & & $\checkmark$ & G8.5 & 3.65 & 3.50 & 3.10\\
Gliese 832  & $\checkmark$ & & M2 & 4.95 & 8.70 & 2.60\\
40 Eri A & & $\checkmark$ & K0 & 4.98 & 4.43 & 4.43\\
70 Oph A & & $\checkmark$ & K0 & 5.08 & 6.00 & 4.75\\
$\alpha$ Aqu & & $\checkmark$ & A7 & 5.13 & 0.76 & 7.56\\
$\sigma$ Dra & & $\checkmark$ & G9 & 5.76 & 4.67 & 5.17\\
$\eta$ Cas A & & $\checkmark$ & G0 & 5.95 & 3.44 & 5.83\\
36 Oph A & & $\checkmark$ & K2 & 5.98 & 5.08 & 5.36\\
82 Eri & & $\checkmark$ & G8 & 6.04 & 4.25 & 4.72\\
$\delta$ Pav & & $\checkmark$ & G8 IV & 6.11 & 3.56 & 6.07\\
$\beta$ Hyi & & $\checkmark$ & G2 IV & 7.46 & 2.80 & 7.85\\
$\pi^3$ Ori & & $\checkmark$ & G2 IV & 8.07 & 3.16 & 9.29\\
$\beta$ Gem & $\checkmark$ & & K0 III & 10.3 & 1.14 & 16.6 \\
55 Cnc A & $\checkmark$ & & G8 & 12.3 & 5.95 & 11.8\\
$\upsilon$ And & $\checkmark$ & & F8 & 13.5 & 4.09 & 15.9\\
47 UMa & $\checkmark$ & & G1 & 14.1 & 5.03 & 14.8\\
$\mu$ Ara  & $\checkmark$ & & G3 & 15.5 & 5.12 & 16.5\\
Gliese 777 & $\checkmark$ & & G6 IV & 15.9 & 5.71 & 14.8\\
14 Her & $\checkmark$ & & K0 & 17.6 & 6.67 & 16.4\\
HD 87883 & $\checkmark$ & & K0 & 18.2 & 7.56 & 15.9\\
HD 39091 & $\checkmark$ & & G1 IV & 18.3 & 5.67 & 19.6\\
HD 154345 & $\checkmark$ & & G8 & 18.6 & 6.74 & 17.1\\
HD 217107 & $\checkmark$ & & G8 IV & 19.9 & 6.17 & 19.4\\
HD 114613 & $\checkmark$ & & G4 IV & 20.7 & 4.85 & 24.0\\
$\psi$ Dra B & $\checkmark$ & & F8 & 23.1 & 5.82 & 25.9\\
HD 142 & $\checkmark$ & & F7 & 25.7 & 5.71 & 29.8\\
HD 134987 & $\checkmark$ & & G5 & 26.2 & 6.45 & 27.0\\
\hline
\end{tabular}
\end{table}

\section{Conclusion}
\label{sec:conclusion}

We constructed an analytical model of the \wfirst\ diffraction spikes generated by a single support strut obscuring the telescope aperture.
This model is a very reasonable description of the actual PSF outside of the first diffraction minimum.
By propagating Poisson noise from the star, the sky, and the thermal emission of the telescope, as well as read-out noise and dark current, we determine that centering on the diffraction spikes of \wfirst\ allows for an astrometric precision of 10\,$\mu$as for a $R=6$ of $J=5$ star in a 100\,s exposure.
The best attainable precision at fixed magnitude is realized in the bluest filters because the diffraction spikes are narrowest.

Given that both the diffraction spike measurement and the systematic contributions from optical distortions and pixel-level artifacts yield diminishing returns for longer integrations, more precise astrometry can be achieved with a series of exposures with $t\lesssim100$\,s.
To better determine the optical distortion pattern and to account for possible spatial correlations of pixel-level artifacts, these exposures should be offset by about $100$ pixels or more, which is the range in which most of the information about the stellar center is contained.
We find that with an assumed systematic uncertainty of 20\,$\mu$as per exposure, a per-visit precision of better than 10\,$\mu$as can be achieved for $R<7.5$ or $J<6.5$ stars with total visit times of 1000\,s or less.

For such bright stars the measurement is limited by systematics. 
Uncorrected optical distortions or small-scale flaws e.g. from subpixel QE variations or persistence can quickly dominate the overall precision of the measurement.
It will be critical for reproducible precision astrometry that the optical distortion model is well constrained at least in the central regions of the detectors.
We believe that these challenges can be met given \wfirst's thermally stable environment in the L2 orbit and the calibration products from the microlensing program and dedicated calibration campaigns in high-stellar density fields.

An astrometric exoplanet discovery program with \wfirst\ could detect Earth-mass planets with orbital periods of 1\,yr or more as well as Neptune-like planets with shorter periods around bright starts within $\sim10\,$pc.
Combining with measurements from \gaia\ could additionally provide access to the regime of rocky planets with periods of 10\,yr or longer.
An astrometric observing program would complement the \wfirst\ direct imaging program and provide masses and orbits for planets  whose atmospheres are characterized with either a starshade or coronagraph.

\section*{Acknowlegements}
PM wants to thank Andrea Bellini, Stefano Casertano, Craig Loomis, and Jim Gunn for instructive discussions.  This research has been supported by the NASA \wfirst\ program.
The Flatiron Institute is supported by the Simons Foundation.

\software{WebbPSF \citep{Perrin2012,Perrin2014}}

\bibliography{references.bib}

\begin{thebibliography}{}
\expandafter\ifx\csname natexlab\endcsname\relax\def\natexlab#1{#1}\fi
\providecommand{\url}[1]{\href{#1}{#1}}

\bibitem[{{Anderson} \& {Francis}(2012)}]{Anderson2012}
{Anderson}, E., \& {Francis}, C. 2012, Astronomy Letters, 38, 331

\bibitem[{{Anderson} \& {King}(2003)}]{Anderson2003}
{Anderson}, J., \& {King}, I.~R. 2003, \pasp, 115, 113

\bibitem[{{Barron} {et~al.}(2007){Barron}, {Borysow}, {Beyerlein}, {Brown},
  {Lorenzon}, {Schubnell}, {Tarl{\'e}}, {Tomasch}, \&
  {Weaverdyck}}]{Barron2007}
{Barron}, N., {Borysow}, M., {Beyerlein}, K., {et~al.} 2007, \pasp, 119, 466

\bibitem[{{Bellini} {et~al.}(2011){Bellini}, {Anderson}, \&
  {Bedin}}]{Bellini2011}
{Bellini}, A., {Anderson}, J., \& {Bedin}, L.~R. 2011, \pasp, 123, 622

\bibitem[{{Blanton} {et~al.}(2005){Blanton}, {Schlegel}, {Strauss},
  {Brinkmann}, {Finkbeiner}, {Fukugita}, {Gunn}, {Hogg}, {Ivezi{\'c}}, {Knapp},
  {Lupton}, {Munn}, {Schneider}, {Tegmark}, \& {Zehavi}}]{Blanton2005}
{Blanton}, M.~R., {Schlegel}, D.~J., {Strauss}, M.~A., {et~al.} 2005, \aj, 129,
  2562

\bibitem[{{Casertano} {et~al.}(2008){Casertano}, {Lattanzi}, {Sozzetti},
  {Spagna}, {Jancart}, {Morbidelli}, {Pannunzio}, {Pourbaix}, \&
  {Queloz}}]{Casertano2008}
{Casertano}, S., {Lattanzi}, M.~G., {Sozzetti}, A., {et~al.} 2008, \aap, 482,
  699

\bibitem[{{Frei} \& {Gunn}(1994)}]{Frei1994}
{Frei}, Z., \& {Gunn}, J.~E. 1994, \aj, 108, 1476

\bibitem[{{Gaia Collaboration} {et~al.}(2016){Gaia Collaboration}, {Prusti},
  {de Bruijne}, {Brown}, {Vallenari}, {Babusiaux}, {Bailer-Jones}, {Bastian},
  {Biermann}, {Evans}, \& et~al.}]{Gaia2016}
{Gaia Collaboration}, {Prusti}, T., {de Bruijne}, J.~H.~J., {et~al.} 2016,
  \aap, 595, A1

\bibitem[{{Gould} {et~al.}(2015){Gould}, {Huber}, {Penny}, \&
  {Stello}}]{Gould2015}
{Gould}, A., {Huber}, D., {Penny}, M., \& {Stello}, D. 2015, Journal of Korean
  Astronomical Society, 48, 93

\bibitem[{{Hardy} {et~al.}(2014){Hardy}, {Willot}, \& {Pazder}}]{Hardy2014}
{Hardy}, T., {Willot}, C., \& {Pazder}, J. 2014, in \procspie, Vol. 9154, High
  Energy, Optical, and Infrared Detectors for Astronomy VI, 91542D

\bibitem[{{Kopparapu} {et~al.}(2014){Kopparapu}, {Ramirez}, {SchottelKotte},
  {Kasting}, {Domagal-Goldman}, \& {Eymet}}]{Kopparapu2014}
{Kopparapu}, R.~K., {Ramirez}, R.~M., {SchottelKotte}, J., {et~al.} 2014,
  \apjl, 787, L29

\bibitem[{{Kopparapu} {et~al.}(2013){Kopparapu}, {Ramirez}, {Kasting}, {Eymet},
  {Robinson}, {Mahadevan}, {Terrien}, {Domagal-Goldman}, {Meadows}, \&
  {Deshpande}}]{Kopparapu2013}
{Kopparapu}, R.~K., {Ramirez}, R., {Kasting}, J.~F., {et~al.} 2013, \apj, 765,
  131

\bibitem[{{Libralato} {et~al.}(2014){Libralato}, {Bellini}, {Bedin}, {Piotto},
  {Platais}, {Kissler-Patig}, \& {Milone}}]{Libralato2014}
{Libralato}, M., {Bellini}, A., {Bedin}, L.~R., {et~al.} 2014, \aap, 563, A80

\bibitem[{{Perrin} {et~al.}(2014){Perrin}, {Sivaramakrishnan}, {Lajoie},
  {Elliott}, {Pueyo}, {Ravindranath}, \& {Albert}}]{Perrin2014}
{Perrin}, M.~D., {Sivaramakrishnan}, A., {Lajoie}, C.-P., {et~al.} 2014, in
  \procspie, Vol. 9143, Space Telescopes and Instrumentation 2014: Optical,
  Infrared, and Millimeter Wave, 91433X

\bibitem[{{Perrin} {et~al.}(2012){Perrin}, {Soummer}, {Elliott}, {Lallo}, \&
  {Sivaramakrishnan}}]{Perrin2012}
{Perrin}, M.~D., {Soummer}, R., {Elliott}, E.~M., {Lallo}, M.~D., \&
  {Sivaramakrishnan}, A. 2012, in \procspie, Vol. 8442, Space Telescopes and
  Instrumentation 2012: Optical, Infrared, and Millimeter Wave, 84423D

\bibitem[{{Perryman} {et~al.}(2014){Perryman}, {Hartman}, {Bakos}, \&
  {Lindegren}}]{Perryman2014}
{Perryman}, M., {Hartman}, J., {Bakos}, G.~{\'A}., \& {Lindegren}, L. 2014,
  \apj, 797, 14

\bibitem[{{Plazas} {et~al.}(2017){Plazas}, {Shapiro}, {Smith}, {Rhodes}, \&
  {Huff}}]{Plazas2017}
{Plazas}, A.~A., {Shapiro}, C., {Smith}, R., {Rhodes}, J., \& {Huff}, E. 2017,
  ArXiv e-prints, arXiv:1703.08205

\bibitem[{{Riess} {et~al.}(2014){Riess}, {Casertano}, {Anderson}, {MacKenty},
  \& {Filippenko}}]{Riess2014}
{Riess}, A.~G., {Casertano}, S., {Anderson}, J., {MacKenty}, J., \&
  {Filippenko}, A.~V. 2014, \apj, 785, 161

\bibitem[{{Roberge} {et~al.}(2015){Roberge}, {Seager}, {Thomson}, {Turnbull},
  {Sparks}, {Shaklan}, {Kuchner}, {Kasdin}, {Domagal-Goldman}, \&
  {Cash}}]{Roberge2015}
{Roberge}, A., {Seager}, S., {Thomson}, M., {et~al.} 2015, in Pathways Towards
  Habitable Planets, 92

\bibitem[{{Seager}(2010)}]{Seager2010}
{Seager}, S. 2010, {Exoplanets}

\bibitem[{{Sozzetti}(2015)}]{Sozzetti2015}
{Sozzetti}, A. 2015, ArXiv e-prints, arXiv:1502.03575

\bibitem[{{Spergel} {et~al.}(2013){Spergel}, {Gehrels}, {Breckinridge},
  {Donahue}, {Dressler}, {Gaudi}, {Greene}, {Guyon}, {Hirata}, {Kalirai},
  {Kasdin}, {Moos}, {Perlmutter}, {Postman}, {Rauscher}, {Rhodes}, {Wang},
  {Weinberg}, {Centrella}, {Traub}, {Baltay}, {Colbert}, {Bennett},
  {Kiessling}, {Macintosh}, {Merten}, {Mortonson}, {Penny}, {Rozo},
  {Savransky}, {Stapelfeldt}, {Zu}, {Baker}, {Cheng}, {Content}, {Dooley},
  {Foote}, {Goullioud}, {Grady}, {Jackson}, {Kruk}, {Levine}, {Melton},
  {Peddie}, {Ruffa}, \& {Shaklan}}]{Spergel2013}
{Spergel}, D., {Gehrels}, N., {Breckinridge}, J., {et~al.} 2013, ArXiv
  e-prints, arXiv:1305.5422

\bibitem[{{Spergel} {et~al.}(2015){Spergel}, {Gehrels}, {Baltay}, {Bennett},
  {Breckinridge}, {Donahue}, {Dressler}, {Gaudi}, {Greene}, {Guyon}, {Hirata},
  {Kalirai}, {Kasdin}, {Macintosh}, {Moos}, {Perlmutter}, {Postman},
  {Rauscher}, {Rhodes}, {Wang}, {Weinberg}, {Benford}, {Hudson}, {Jeong},
  {Mellier}, {Traub}, {Yamada}, {Capak}, {Colbert}, {Masters}, {Penny},
  {Savransky}, {Stern}, {Zimmerman}, {Barry}, {Bartusek}, {Carpenter}, {Cheng},
  {Content}, {Dekens}, {Demers}, {Grady}, {Jackson}, {Kuan}, {Kruk}, {Melton},
  {Nemati}, {Parvin}, {Poberezhskiy}, {Peddie}, {Ruffa}, {Wallace}, {Whipple},
  {Wollack}, \& {Zhao}}]{Spergel2015}
{Spergel}, D., {Gehrels}, N., {Baltay}, C., {et~al.} 2015, ArXiv e-prints,
  arXiv:1503.03757

\end{thebibliography}

\appendix

\section{Pixel-integrated PSF of a rectangular obstruction, and its derivatives}
\label{sec:derivation}

Realizing that both the PSF intensity in \autoref{eq:psf} as well as the pixel shape in \autoref{eq:psfp}  are separable into two 1D functions, we can break down the entire integral into 1D.
We utilize the convolution theorem and these known analytic results for Fourier transforms:
\begin{align}
&\mathrm{FT}\left[f(x)\right](q) \equiv \frac{1}{\sqrt{2\pi}}\int dx\ f(x) \mathrm{e}^{-iqx}\\
&\mathrm{FT}\left[\text{rect}(cx)\right](q) = \frac{1}{\sqrt{2\pi c^2}}\ \text{sinc}\left(\frac{q}{2 c}\right)\\
&\mathrm{FT}\left[\text{sinc}^2(cx)\right](q) = \frac{2}{\sqrt{2\pi c^2}}\ \text{tri}\left(\frac{q}{2 c}\right),
\end{align} 
where $\text{tri}(x)\equiv\text{rect}(x/2)(1-|x|)$. 
We can now express the $x$-part of \autoref{eq:psfp} as follows:
\begin{equation}
\mathrm{FT}\left[I_p(x)\right] = \frac{2}{\sqrt{2\pi} k_x}\ \text{tri}\left(\frac{q}{2 k_x}\right) \frac{w}{\sqrt{2\pi}}\ \text{sinc}\left(\frac{q w}{2}\right).
\end{equation}
The inverse Fourier transform of that equation is
\begin{equation}
\label{eq:psf1d}
\begin{split}
I_p(x\,|\,k,w) =  \frac{1}{k \pi (w^2 - 4x^2)}\Big[&
\big[-2w+2w\cos(kw)\cos(2 kx) + 4 x \sin(kw) \sin(2 kx) +\\
&k(w^2-4 x^2) \left[\text{Si}\big(k(w-2 x)\big) + \text{Si}\big(k(w+2 x)\big) \right]
\Big],
\end{split}
\end{equation}
where $\text{Si}$ denotes the sine integral and we already applied a normalization of $\sqrt{\tfrac{2}{\pi}}\frac{w}{k}$ so the the function integrates to unity.

The derivatives of this function are:
\begin{equation}
\label{eq:dpsf1d}
\frac{\partial I_p(x\,|\,k,w)}{\partial x} = 
\frac{16 w  x \left(-1 +\cos(kw)\cos(2kx)\right) + 4 (w^2+4 x^2) \sin(k w) \sin(2kx)}{k \pi \left(w^2-4 x^2\right)^2}
\end{equation}
\begin{equation}
\begin{split}
\frac{\partial^2 I_p(x\,|\,k,w)}{\partial x^2} = \frac{8}{k \pi \left(w^2-4 x^2\right)^3}
\Big[&-2w (w^2+12x^2)\ +\\
&\cos(2 k x)\left(2w (w^2+12x^2)\cos(k w) + k(w^4-16x^4)\sin(k w)\right) +\\
&4x\left(-kw (w^2-4x^2)\cos(k w)+(3w^2+4x^2)\sin(k w)\right)\sin(2k x)\Big]
\end{split}
\end{equation}

\end{document}